\documentclass[aps,prc,twocolumn,superscriptaddress,showpacs]{revtex4}

\usepackage{graphicx}
\usepackage{times,mathptmx}
\usepackage{amsmath}
\usepackage{amssymb}
\usepackage{dcolumn}

\begin{document}

 \title{Systematics of threshold incident energy for 
 deep sub-barrier fusion hindrance} 

\author{Takatoshi Ichikawa}%
\affiliation{RIKEN, Wako, Saitama 351-0198, Japan}
\author{Kouichi Hagino}
\affiliation{Department of Physics, Tohoku University, Sendai 980-8578, Japan}
\author{Akira Iwamoto}
\affiliation{Japan Atomic Energy Agency, Tokai-mura, Naka-gun, Ibaraki
319-1195, Japan}

\date{\today}

\begin{abstract}
We systematically evaluate the potential energy at the touching 
configuration for heavy-ion reactions using various potential models. 
We point out that the energy at the touching point, especially that 
estimated with the Krappe-Nix-Sierk (KNS) potential, strongly
correlates with the threshold incident energy for steep fall-off
of fusion cross sections observed recently for several systems 
at extremely low energies.
This clearly indicates that 
the steep fall-off phenomenon can be attributed to 
the dynamics after the target and
projectile touch with each other, e.g., the 
tunneling process 
and the nuclear saturation property in the overlap region. 
\end{abstract}


\maketitle
Recently, for medium-heavy mass systems, it has become possible 
to measure fusion cross sections down to extremely low
incident energies.
In those measurements, unexpected steep fall-off of fusion 
cross sections, as compared to a standard theoretical calculation, 
have been observed at deep sub-barrier
energies
~\cite{PhysRevLett.89.052701,jiang04,jiang:014604,jiang:044613,jiang06,2006PhLB..640...18J}.  
Although 
the steep fall-off phenomenon, referred to as the fusion hindrance, 
may be accounted for if one uses 
an anomalously large diffuseness parameter in the Woods-Saxon 
potential~\cite{hagino:054603}, 
the physical origin of the phenomenon has yet to be
clarified~\cite{DHLN06}.

\begin{figure}
\includegraphics[keepaspectratio,width=\linewidth]{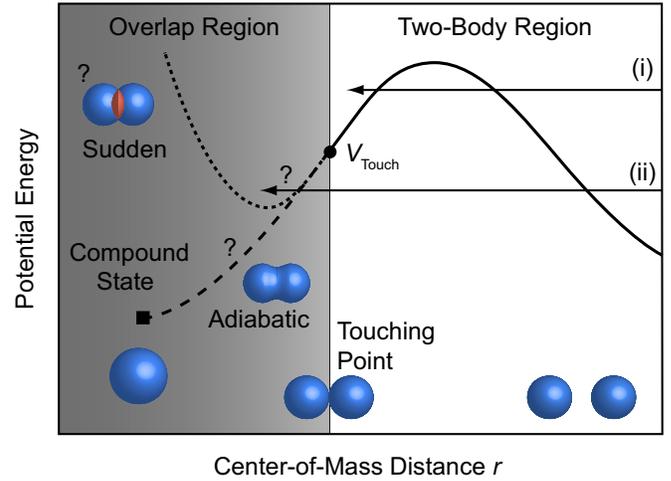}%
\caption{\label{fig1} (Color online) Schematic picture for heavy-ion 
sub-barrier fusion reactions. The filled circle denotes the energy at the
touching point, $V_{\rm Touch}$.
}
\end{figure}

One important aspect of fusion reactions at deep subbarrier energies 
is that the inner turning point of the potential 
may be located far inside the touching point of the colliding nuclei. 
We show this schematically in Fig. 1. 
At energies close to the Coulomb barrier, 
the inner turning point is 
still far outside of the touching point~\cite{DHRS98} 
(see the line (i) in Fig.~\ref{fig1}). 
At these energies, one usually assumes that 
a compound nucleus is automatically formed 
once the projectile penetrates the Coulomb
barrier, 
due to the strong nuclear attractive force in the classically 
allowed region. 
In contrast, 
at energies below the potential energy at the
touching point, $V_{\rm Touch}$, the inner turning point appears more inside
of the touching point (see the line (ii) in Fig.~\ref{fig1}).
That is, the projectile nucleus is still in the classically 
forbidden region when the two colliding nuclei 
touch with each other. 
After the touching, 
an elongated composite system is formed, which evolves in the 
classically forbidden region towards a compound nucleus 
by overlapping between the projectile-like and the 
target-like fragments. 
Since this involves the 
penetration of the residual Coulomb barrier, 
naturally the fusion cross sections are hindered by the tunneling
factor. 

In this paper, 
we evaluate the potential energy at the touching 
configuration for several systems,  
and investigate whether the dynamics after the touching point is 
responsible for the steep fall-off phenomenon. 
In this respect, it is interesting to notice that 
the authors of Refs.~\cite{PhysRevLett.89.052701,jiang:044613} have 
argued that the steep fall-off phenomenon 
systematically takes place below a certain threshold incident
energy, $E_s$. 
We will show below that there is a strong correlation 
between the touching energy $V_{\rm Touch}$ and the threshold 
energy $E_s$, indicating that 
the density overlap in the classically forbidden 
region indeed plays an important
role. We mention that 
one would have to 
settle a model in the overlap region, such as the adiabatic or sudden
models, or some combination of these two, in order to clarify the
whole dynamics of deep subbarrier fusion reactions. 
However, our analysis is independent of these modellings, 
since both the adiabatic and the sudden approaches provide a similar 
potential energy to each other as long as the touching point is 
concerned. 

\begin{table*}
\caption{\label{tab1}
Potential energy at the touching configuration 
calculated by various theoretical
 models for the systems discussed in Ref.~\cite{jiang06}. 
In the second column, $Z_TZ_P$ is the charge product of
the system, while $\mu$ is the reduced mass.  
$X_{\rm eff}$ is the effective fissility parameter defined as 
$X_{\rm eff}=Z_TZ_P/[A_T^{1/3}A_P^{1/3}(A_P^{1/3}+A_T^{1/3})]/12.6$, while 
$E_s$ is the experimental energy 
at which the astrophysical S-factor has 
the maximum~\cite{jiang06}. 
 $V_{\rm KNS}$, $V_{\rm Prox}$, $V_{\rm Bass}$ and 
 $V_{\rm AW}$ denote the results of the Krappe-Nix-Sierk, the proximity,
 the Bass and the Aky\"uz-Winther models, respectively.}
\begin{ruledtabular}
\begin{tabular}{cccccccc}

\multicolumn{1}{c}{System} &
\multicolumn{1}{c}{$Z_TZ_P\sqrt{\mu}$} &
\multicolumn{1}{c}{$X_{\rm eff}$} &
\multicolumn{1}{c}{$E_s$} &
\multicolumn{1}{c}{$V_{\rm KNS}$} &
\multicolumn{1}{c}{$V_{\rm Prox}$} &
\multicolumn{1}{c}{$V_{\rm Bass}$} &
\multicolumn{1}{c}{$V_{\rm AW}$} \\
\multicolumn{1}{c}{}&
\multicolumn{1}{c}{(MeV$^{1/2}/c$)}&
\multicolumn{1}{c}{}&
\multicolumn{1}{c}{(MeV)}&
\multicolumn{1}{c}{(MeV)}&
\multicolumn{1}{c}{(MeV)}&
\multicolumn{1}{c}{(MeV)}&
\multicolumn{1}{c}{(MeV)}\\
\hline                                     
(Type I)                 &        &       &       &       &         &   &   \\   
$^{90}$Zr + $^{90 }$Zr &  10733 & 0.705 & 175 $\pm$ 1.8 & 179.9 &   169.6 &   167.6 &   175.2 \\
$^{90}$Zr + $^{89 }$Y  &  10436 & 0.692 & 171 $\pm$ 1.7 & 175.2 &   164.8 &   162.4 &   170.4 \\
$^{90}$Zr + $^{92 }$Zr & 10792  & 0.698 & 171 $\pm$ 1.7 & 179.1 &   168.8 &   166.4 &   174.4 \\
$^{58}$Ni + $^{58 }$Ni &  4222  & 0.536 & 94 $\pm$ 0.9 &  93.4 &   80.8  &    79.2 &    87.5 \\
$^{60}$Ni + $^{89 }$Y  &  6537  & 0.592 & 123 $\pm$ 1.2 & 125.4 &   113.6 &   111.1 &   119.8 \\
$^{32}$S  + $^{89 }$Y  &  3026  & 0.457 & 72.6 $\pm$ 0.7 &  72.2 &   59.7  &    56.7 &    65.4 \\
\hline
(Type II)                &        &       &       &       &         &   &   \\   
$^{64}$Ni + $^{100}$Mo &  7343  & 0.582 & 121 $\pm$ 1.2 & 131.7 &   120.0 &   115.9 &   126.2 \\
$^{64}$Ni + $^{ 64}$Ni &  4435  & 0.486 & 87.3 $\pm$ 0.9 &  89.0 &   76.1  &    71.9 &    82.9 \\
\hline
(Type III)                 &        &       &       &       &         &   &   \\   
$^{48}$Ca + $^{ 48}$Ca &  1960  & 0.331 & 48.1 $\pm$ 0.9 &  42.2 &   27.7  &    21.9 &    35.4 \\
$^{28}$Si + $^{ 64}$Ni &  1729  & 0.364 & 47.3 $\pm$ 0.9 &  43.9 &   30.5  &    27.1 &    36.7 \\
$^{16}$O  + $^{ 76}$Ge & 930.5  & 0.282 & 27.6 $\pm$ 0.8 &  26.1 &   13.1  &     9.6 &    18.3 \\
\hline
(Type IV)                 &        &       &       &       &         &   &   \\   
$^{16}$O  + $^{ 16}$O  & 181.0  & 0.159 & 7.1 $\pm$ 0.8 &   2.2 &  -11.4  &   -13.4 &    -5.4 \\
$^{12}$O  + $^{ 16}$O  & 125.7  & 0.137 & $<6.2$ &   0.2 &  -13.2  &   -14.8 &    -7.4 \\
$^{12}$O  + $^{ 14}$N  & 106.8  & 0.129 & $<5.0$ &  -0.5 &  -13.9  &   -15.4 &    -8.1 \\
$^{12}$O  + $^{ 13}$C  &  89.9  & 0.114 & $<4.0$ &  -1.5 &  -14.9  &   -16.6 &    -9.3 \\
$^{11}$O  + $^{ 12}$C  &  71.9  & 0.104 & $<3.0$ &  -2.2 &  -15.5  &   -16.9 &    -9.9 \\
$^{10}$B  + $^{ 10}$B  &  55.9  & 0.099 & $<1.9$ &  -2.2 &  -15.3  &   -16.0 &    -9.9 \\
				        
\end{tabular}
\end{ruledtabular}
\end{table*}  

\begin{figure}
\includegraphics[keepaspectratio,width=\linewidth]{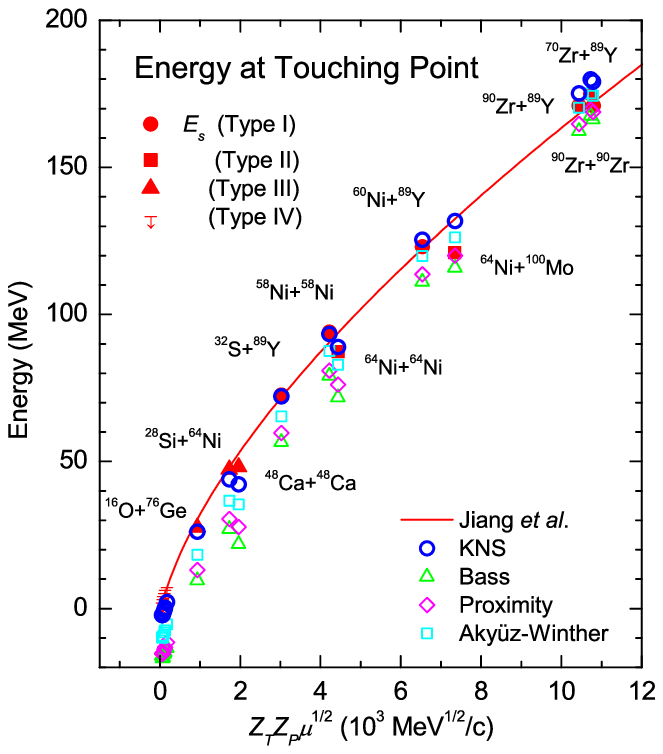}%
\caption{\label{fig2} (Color online) Potential energy at the touching point
 calculated by various theoretical models.
The open circles, triangles, diamonds and squares denote the results of the
 Krappe-Nix-Sierk, the Bass, the Proximity and the Aky\"uz-Winther
 models, respectively.
The solid line denotes the systematics proposed by Jiang {\it et al.} 
~\cite{jiang06}. The filled circles, squares, triangles, and the 
horizontal lines show the experimental energy taken from 
Ref.~\cite{jiang06} 
at which 
the astrophysical S-factor has the maximum value. 
}
\end{figure} 

In order to estimate the potential energy at the touching point, 
$r_{\rm touch}=R_P+R_T$, 
we employ the Krappe-Nix-Sierk
(KNS)~\cite{kr79}, the Bass~\cite{bass1980nrh}, 
the proximity~\cite{PhysRevC.62.044610} and the Aky\"uz-Winther
(AW)~\cite{AW91} models. 
Assuming the spherical shape for both the projectile and target
nuclei, the KNS potential energy at the touching point reads
\begin{eqnarray}
V^{(N)}_{\rm KNS} = -D \left
(4+\frac{r_{\rm touch}}{a}-\frac{f(R_T/a)}{g(R_T/a)}
-\frac{f(R_P/a)}{g(R_P/a)}\right),
\end{eqnarray}
where the functions $f$ and $g$ are defined as $f(x)=x^2\sinh(x)$ and
$g(x)=x\cosh(x)-\sinh(x)$, respectively.
In this model, 
the nuclear radius is given by $R=r_{0}A^{1/3}$,
and the depth constant $D$ by 
\begin{equation}
D=\frac{4
\sqrt{c_{s}^{(T)}c_{s}^{(P)}}a^3}{r_{0}^2\,r_{\rm touch}}\,
g(R_T/a)\,g(R_P/a)e^{-r_{\rm touch}/a},
\end{equation}
where the effective surface energy constant $c_{s}$ is given by
$c_{s}=a_{s}(1-\kappa_{s}I^2)$ with $I=(N-Z)/A$.
We take the parameters to be $a=0.68$ fm,
$a_{s}=21.33$ MeV and $\kappa_{s}=2.378$ from
FRLDM2002~\cite{PhysRevC.62.044610}, 
except for the radius
parameter for which we slightly adjust to be $r_{0}=1.2$ fm 
in order to fit the experimental fusion cross sections for 
the $^{64}$Ni+$^{64}$Ni reaction at energies above the Coulomb barrier. 

The proximity potential energy at the touching point is given by
\begin{eqnarray}
V^{(N)}_{\rm Prox}=-1.7818\frac{b\overline{R}\,a_{2}}{r_{0}^2}-3.00,
\label{prox}
\end{eqnarray}
where $\overline{R}=R_TR_P/(R_T+R_P)$. 
In this model, the nuclear radius 
is given by 
\begin{equation}
R=
R_{00}\left(1-\frac{7}{2}\frac{b^2}{R_{00}^2}-\frac{49}{8}\frac{b^4}{R_{00}^4}
\right)
+\frac{N}{A}\,t
\end{equation}
with 
\begin{eqnarray}
R_{00}&=&1.240A^{1/3}\left[1+1.646/A-0.191(A-2Z)/A)\right], \\
t&=&\frac{3}{2}r_{0}\cdot
\frac{J(N-Z)/A-\frac{1}{12}c_{1}ZA^{-1/3}}{Q+\frac{9}{4}JA^{-1/3}}.
\end{eqnarray}
The value of the parameters are taken to be 
$b$=1 fm, $r_{0}=1.14$ fm, 
$J$=32.65 MeV, $c_{1}$=0.757895 MeV and 
$Q$=35.4 MeV.
The surface energy coefficient $a_{2}$ in Eq. (\ref{prox}) 
is given by $a_{2}=18.36-Q(t_T^2+t_P^2)/2r_{0}^2$.
In order to fit the experimental data, we 
use the same prescription as in Refs.~\cite{sw04,sw05} and 
subtract 
3.00 MeV from the 
original proximity model (the last term in Eq.~(\ref{prox})). 

The Bass potential energy at the touching point is given by 
\begin{eqnarray}
V^{(N)}_{\rm Bass}=-\overline{R}\left[\alpha+\beta \right]^{-1},
\end{eqnarray}
where the parameters $\alpha$ and $\beta$ are taken as $\alpha$=0.0300
MeV$^{-1}$ fm and $\beta$=0.0061 MeV$^{-1}$ fm, respectively. 
In the Bass model, the
nuclear radius is given by $R=1.16 A^{1/3}-1.39A^{-1/3}$. 
The AW potential energy at the touching point, on the other hand,
reads 
\begin{eqnarray}
V^{(N)}_{\rm AW}=-8\pi\gamma \overline{R}a,
\label{AW}
\end{eqnarray}
where the average surface tension $\gamma$ is given by
$\gamma=0.95\left[1-1.8(\frac{N_T-Z_T}{A_T})(\frac{N_P-Z_P}{A_P})\right]$
MeV fm$^{-2}$.   
In this model, 
the nuclear radius is given by $R=(1.20 A^{1/3}-0.09)$ fm
and the diffuseness parameter $a$ is given by 
$a=0.855\cdot[1+0.53(A_T^{-1/3}+A_P^{-1/3})]^{-1}$ fm.

In order to estimate the total potential energy at the touching point, 
$V_{\rm Touch}$, one has to add the Coulomb potential to the nuclear 
potential energies given by Eqs. (\ref{prox}) - (\ref{AW}). 
To this end, we use the Coulomb potential for two point charges, 
$V^{(C)}=e^2Z_TZ_P/r_{\rm touch}$, 
where the touching radius $r_{\rm touch}$ is specified for each 
model for the nuclear potential, $V^{(N)}$. 
The resultant touching energy for the systems discussed in Ref.~\cite{jiang06} is
shown in Fig.~\ref{fig1} as a function of 
$Z_TZ_P\mu^{1/2}$, where $\mu$ is the reduced mass 
of the colliding nuclei. 
The results of the KNS, the Bass, the proximity and
the AW models are denoted by the open circle, the open triangle, the open
diamond and the open square, respectively.
All the results are summarized in Table 1. 
These touching energies are compared with 
the energy $E_s$, at which 
the experimental fusion cross section is maximum when it is plotted 
in terms of the astrophysical S-factor~\cite{jiang06}.     
These ``experimental'' energies $E_s$ are 
shown in Fig. 2 by the filled circles, 
the filled squares, the filled triangles, and the 
horizontal lines, depending on the types of the system as defined in
Ref. \cite{jiang06}. 
Notice that the energy $E_s$ for the type III 
was estimated by extrapolation, and that for the type IV is only an
upper limit. 
The systematics for the energy $E_s$ proposed by 
Jiang {\it et al.}~\cite{jiang06} is also shown by the solid line.

Althogh the physical significance for the energy $E_s$ is not
clear, because the S-factor representation for fusion cross sections 
would be useful only at much lower energies 
than the lowest energies of the current measurements, at which the outer
turning point is much larger than the inner turning point (see e.g., 
Refs. \cite{MS73,SRR07} for a discussion on the modified S-factor, that 
takes into account the effect of the inner turning point), 
it is 
remarkable that the result of the KNS model follows closely to
the energy $E_s$, and thus the systematics shown
by the solid line (an exceptional case of $^{64}$Ni+$^{100}$Mo will be
discussed in the next paragraph). 
The good correspondence between $V_{\rm KNS}$ and $E_s$ may be due to the 
fact that 
the KNS model partly takes into account the 
saturation of nuclear matter when two nuclei come inside the
Coulomb barrier\cite{IHI07} (in fact, the KNS model has been shown to be 
consistent with the energy density formalism
with the Skyrme SkM$^{*}$ interaction~\cite{vaz81,de02}).
The result of the AW potential is similar to that of
the KNS model, although the deviation from $E_s$ is slightly larger. 
For the Bass and the proximity models, although 
the dependence of the touching
energy $V_{\rm Touch}$ on the parameter $Z_TZ_P\mu^{1/2}$
is similar to that of the 
KNS and the AW models, there is a large discrepancy between 
the touching energy $V_{\rm touch}$ and the threshold energy $E_s$. 

\begin{table}
\caption{\label{tab2}
The potential energy at the touching configuration calculated with the
 KNS model for the systems discussed in
 Refs.~\cite{jiang04,jiang:014604,jiang:044613}. 
All of these systems are categolized as Type III.} 
\begin{ruledtabular}
\begin{tabular}{cccccc}

\multicolumn{1}{c}{System} &
\multicolumn{1}{c}{$Z_TZ_P\sqrt{\mu}$} &
\multicolumn{1}{c}{$X_{\rm eff}$} &
\multicolumn{1}{c}{$E_s$} &
\multicolumn{1}{c}{$V_{\rm KNS}$} &
\multicolumn{1}{c}{Ref.}\\
\multicolumn{1}{c}{}&
\multicolumn{1}{c}{(MeV$^{1/2}/c$)}&
\multicolumn{1}{c}{}&
\multicolumn{1}{c}{(MeV)}&
\multicolumn{1}{c}{(MeV)}&
\multicolumn{1}{c}{}\\
\hline
$^{34}$S  + $^{89}$Y  &  3095&      0.444&  72.6  &   70.9&\cite{jiang04}\\
$^{28}$Si  + $^{58}$Ni&  1704&      0.383&  49    &   45.3&\cite{jiang04}\\
$^{28}$Si  + $^{62}$Ni&  1722&      0.370&  48.6  &   44.3&\cite{jiang04}\\
$^{16}$O  + $^{208}$Pb&  2529&      0.413& 69.6   &  70.5 &\cite{jiang:014604}\\
$^{16}$O + $^{144}$Sm &  1882&      0.384&    57.7&   54.6&\cite{jiang:014604}\\
$^{19}$F + $^{208}$Pb &  3079&      0.431&    75.5&   78.8&\cite{jiang:014604}\\
$^{40}$Ca + $^{90}$Zr &  4210&      0.524&    93.2&   93.6&\cite{jiang:014604}\\
$^{50}$Ti + $^{208}$Pb& 11454&      0.683&   181.2&  191.9&\cite{jiang:014604}\\
$^{58}$Ni + $^{60}$Ni &  4258&      0.527&  92 $\pm$ 2  &   92.5&\cite{jiang:044613}\\
$^{58}$Ni  + $^{64}$Ni&  4325&      0.511&  89 $\pm$ 2  &   91.1&\cite{jiang:044613}\\
$^{58}$Ni  + $^{74}$Ge&  5109&      0.542&  98.5 $\pm$ 2.0 &  103.9&\cite{jiang:044613}\\
$^{64}$Ni  + $^{74}$Ge&  5249&      0.517&  97.5 $\pm$ 2.0 &  101.5&\cite{jiang:044613}\\
				        
\end{tabular}
\end{ruledtabular}
\end{table}  

\begin{figure}
\includegraphics[keepaspectratio,width=\linewidth]{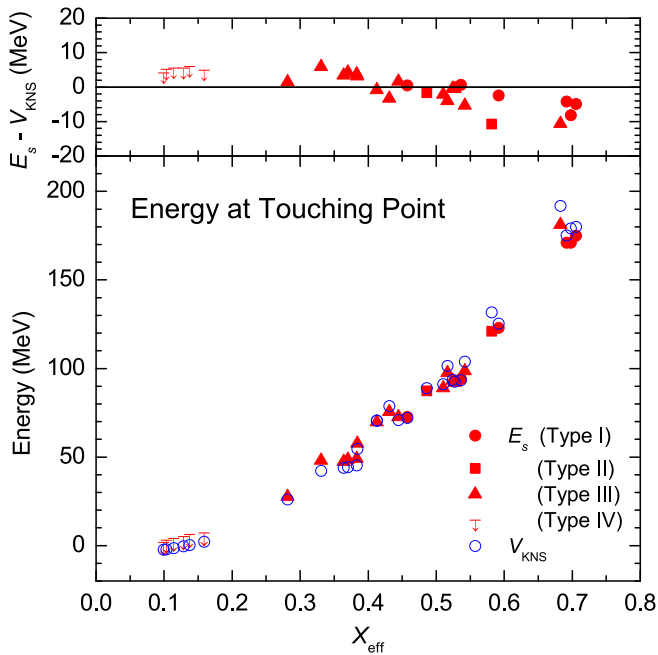}%
\caption{\label{fig3} (Color online) 
Same as Fig. 2, but as a function of the 
effective fissility parameter defined in Ref.~\cite{bass1980nrh}.
The upper panel shows the difference between $E_s$ and $V_{\rm KNS}$.
}
\end{figure} 

For the asymmetric $^{64}$Ni+$^{100}$Mo reaction,
the experimental threshold energy $E_s$ deviates largely from the
systematics curve.  
The calculations with the KNS and AW models are consistent with the
systematics curve but not to the value of $E_s$.  
In order to check how the touching energy $V_{\rm Touch}$ compares
with the threshold energy $E_s$ for other asymmetric systems, 
we also examine the
$^{16}$O+$^{208}$Pb reaction. 
For this system, we find that the KNS model leads to the touching 
energy $V_{\rm Touch}$ that is consistent with
the experimental threshold energy $E_s$ 
~\cite{jiang:014604} (see Table II).
Therefore, it is unlikely that 
the large difference between $V_{\rm Touch}$ and $E_s$ 
for the $^{64}$Ni+$^{100}$Mo system 
can be attributed to the model assumption of the KNS potential. 
Notice that for this system, there may exist
some peculiar nuclear structure effect, 
because 
the coupled-channels 
calculation reported in Ref.~\cite{esbensen05} 
does not seem to account well for the experimental fusion 
cross sections even above the threshold energy $E_s$.  
A further 
investigation is necessary for this system 
concerning the threshold energy. 

In order to see more clearly the correlation between $E_s$ and
$V_{\rm KNS}$, 
the lower panel of Fig.~\ref{fig3} shows these energies 
as a function of the effective fissility parameter $X_{\rm eff}$ 
defined as 
$X_{\rm eff}=Z_TZ_P/[A_T^{1/3}A_P^{1/3}(A_P^{1/3}+A_T^{1/3})]/12.6$~\cite{bass1980nrh}. 
The figure includes also a few more systems than shown in Fig. 2, 
which are taken from
Refs.~\cite{jiang04,jiang:014604,jiang:044613} (see Table II for the
additional data). 
With this representation, 
all the data points distribute
more uniformly than in Fig.~\ref{fig2}.
The difference between $E_s$ and $V_{\rm KNS}$ is also shown in
the upper panel of Fig.~\ref{fig3}.
One observes that 
the difference between $E_s$ and
$V_{\rm KNS}$ is indeed small except for large $X_{\rm eff}$, 
clearly indicating that there is a strong correlation 
between these two values.
The large discrepancy for systems with large
$X_{\rm eff}$ may be due to the ambiguity of the experimental data,
because the measurements were only for the fusion-evaporation cross
sections and the fusion-fission cross sections were estimated 
using the statistical model~\cite{jiang04,jiang:014604,jiang:044613}.

In summary, we have shown that the potential energy 
at the touching point
strongly correlates with the threshold incident energies 
for the steep fall-off of the fusion cross sections.
The systematics of the threshold energy can be rather naturally
explained by the present approach in terms of the touching energy.  
This strongly suggests that 
the overlap process after the touching 
is responsible for the steep fall-off of the fusion
cross section.
For such overlap process, the sudden and adiabatic approaches have been
often employed~\cite{bal98,HW06}. 
In the former, the frozen density approximation while overlapping with
the colliding nuclei is applied, and in the latter, the dynamical change
in the density of the colliding nuclei is taken into account.
These two approaches are in the opposite limit to each other, and 
there is not yet a definite consensus regarding 
which limit better describes the 
realistic situation at deep subbarrier energies. 
In this respect, 
the threshold energy discussed in this paper 
will provide a useful constraint to modelling of the overlap
process as the touching configuration is a doorway of such process. 

\begin{acknowledgments}
This work was supported by the Grant-in-Aid for Scientific Research,
Contract No. 19740115 from the Japanese Ministry of Education,
Culture, Sports, Science, and Technology.
\end{acknowledgments}                     


\end{document}